\documentclass[12pt,a4paper]{article}
\usepackage{jcappub}
\usepackage{euscript,epsfig,amsmath,amssymb}
\usepackage{amsfonts,latexsym}
\usepackage[normalem]{ulem}
\usepackage{color}

\newcommand{\be}[1]{\begin{equation}\label{#1}}
\newcommand{\ee}{\end{equation}}
\newcommand{\ba}[1]{\begin{eqnarray}\label{#1}}
\newcommand{\ea}{\end{eqnarray}}
\newcommand{\rf}[1]{(\ref{#1})}
\newcommand{\nn}{\nonumber}

\newcommand{\ov}{\overline}

\begin{document}

\title{Cosmic screening of the gravitational interaction}

\author[a]{Maxim Eingorn,}
\author[b]{Claus Kiefer}
\author[c,d
]{and Alexander Zhuk
}

\affiliation[a]{North Carolina Central University, CREST and NASA Research Centers,\\ 1801 Fayetteville st., Durham, North Carolina 27707, U.S.A.\\}

\affiliation[b]{Institute for Theoretical Physics, University of Cologne,\\ Z\"ulpicher Stra\ss e 77, 50937 K\"oln, Germany\\}

\affiliation[c]{The International Center of Future Science of the Jilin University,\\ 2699 Qianjin st., 130012, Changchun City, China\\}

\affiliation[d]{Astronomical Observatory, Odessa National University,\\ Dvoryanskaya st. 2, Odessa 65082, Ukraine\\}

\emailAdd{maxim.eingorn@gmail.com} \emailAdd{kiefer@thp.uni-koeln.de} \emailAdd{ai.zhuk2@gmail.com}

\abstract{We study a universe filled with cold dark matter in the form of discrete inhomogeneities (e.g., galaxies) and dark energy in the form of a continuous
perfect fluid. We develop a first-order scalar perturbation theory in the weak gravity limit around a spatially flat Friedmann universe. Our approach works at all
cosmic scales and incorporates linear and nonlinear effects with respect to energy density fluctuations.  The gravitational potential can be split into individual
contributions from each matter source. Each potential is characterized by a Yukawa interaction with the same range, which is of the order of 3700 Mpc at the
present time. The derived equations can form the theoretical basis for numerical simulations for a wide class of modern cosmological models.

\

\noindent {\it This essay was selected for Honorable Mention in the 2017 Gravity Research Foundation competition.}}


\maketitle

\flushbottom


In 1692, the young clergyman Richard Bentley had started a correspondence with Isaac Newton, in which they discussed, among other things, the question whether the
universe is unbounded or not. In a first response to Bentley's queries, Newton expressed the opinion that the universe must necessarily be unbounded, because
otherwise the attractive gravitational force would lead to a collapse of all matter into the center. Newton then continued to write (see \cite{Harrison}, p.~60):
\begin{quote}
But if the matter was evenly diffused through an infinite space, it would never convene into one mass but some of it into one mass and some into another so as to
make an infinite number of great masses scattered at great distances from one to another throughout all of infinite space. And thus might the Sun and first stars
be formed.
\end{quote}
An unbounded universe leads, however, to infinite attraction forces, so in Newtonian theory the universe must be exactly isotropic to avoid collapse. As Newton
wrote in the second edition of his {\em Principia} (published after the Bentley letters) (see \cite{Harrison}, p.~61):
\begin{quote}
The fixed stars, being equally spread out in all points of the heavens, cancel out their mutual pulls by opposite attractions.
\end{quote}
Such an exact isotropy is, of course, puzzling. In the 19th century, scientists thought about alternatives, and one alternative was a modification of Newton's law
at large scales \cite{Norton}.
For example, in 1895, the German astronomer Hugo von Seeliger envisaged a modification by what we now call a ``Yukawa force''; in
\cite{Seeliger}, equation~(2), he suggested a form for the gravitational attraction force given by $k^2mm'{\rm e}^{-\lambda r}/r^2$ (from an ``absorption effect''
of space, as he called it). As there was a firm belief in the static nature of the universe as a whole, no explanation was sought in terms of a dynamically
evolving universe.

This Newtonian problem was eventually solved by Einstein's theory of general relativity, in which gravity is no longer a force acting at a distance, but a
manifestation of the local geometry of spacetime. In order to get a static universe, Einstein introduced in 1917 the cosmological constant, but it was soon
recognized that his static solution is unstable and that his equations predict an expanding (or collapsing) universe.

In spite of this, it is still of great interest to study the exact modification of the Newtonian forces at cosmic scales {\em as predicted from general
relativity}. This is what we will do in our essay. We will show, in particular, that the gravitational attraction at large scales is subject to a Yukawa-type
modification and we will give a concrete number for its range. This does not only shed new light on the old problem discussed in the Newton-Bentley
correspondence, but also gives a prediction for the largest observable structures in the universe and serves as a starting point for numerical investigations of
structure formation.

To be specific, we consider two types of ``matter'' sources. The first type is discrete inhomogeneities (of masses $m_n$), such as galaxies, and mostly represents
cold dark matter (CDM). This source is characterized by the rest mass density $\rho_{\rm M}\,=\,\sum_n m_n \delta({\bf r}-{\bf r}_n) \equiv \sum_n \rho_n$. The
second type is a set of an arbitrary number of continuous fluids with barotropic equations of state; it can represent, in particular, a dynamical dark energy (DE)
as well as radiation. We consider a spatially flat Friedmann universe with small metric perturbations, but we impose no restriction on the energy densities.

Let us first address the evolution of the background, which is given by the Friedmann-Lema\^{\i}tre equations (see, e.g., \cite{Mukhanov-book})
\be{2} \frac{3{\mathcal
    H}^2}{a^2}=\frac{3H^2}{c^2}=\kappa\left(\overline{\varepsilon}_{\rm
    M}+\sum\limits_I\overline{\varepsilon}_I\right)\, ,\ee
\be{3} \frac{2{\mathcal H}'+{\mathcal H}^2}{a^2}= \frac{1}{c^2}\left(3H^2+2\dot H\right)=\frac{H^2}{c^2}\left(1-2q\right)= -\kappa\sum\limits_I\overline{p}_I\, ,
\ee
with the equations of state $\; \ov p_{\rm M}=0, \quad \ov p_I=\omega_I\ov \varepsilon_I,\, \, \omega_I=\mathrm{const}\neq0$ (barred quantities correspond to the
background).

In these equations, $H\equiv \dot a/a\equiv (da/d t)/a$ is the standard Hubble parameter and $\mathcal H\equiv a'/a\equiv (da/d\eta)/a$ is the Hubble parameter
with respect to conformal time $\eta$; we have $c{\mathcal H}=aH$. Moreover, $q\equiv-(\ddot a/a)/H^2$ is the deceleration parameter. Primes and overdots denote
derivatives with respect to conformal and standard (synchronous) time, respectively. We have $\kappa\equiv 8\pi G_{\rm N}/c^4$, where $G_{\rm N}$ is Newton's
constant and $c$ is the speed of light.  The case $\omega=-1$ corresponds to the cosmological constant (i.e. $\kappa \ov\varepsilon_{\Lambda} \equiv \Lambda$).
Since the $\Lambda$-term only affects the background and has no perturbations, we will assume that $\omega\neq -1$ in equations related to the perturbation
theory. Obviously, $\ov{\varepsilon}_{\rm M}= \ov{\rho}_{\rm M}c^2/a^3$, $\ov{\rho}_{\rm M}=\mathrm{const}$.

Equations \rf{2} and \rf{3} lead to a useful relation
\be{5} \frac{3}{2}\kappa\left[\overline{\varepsilon}_{\rm M}+\sum\limits_I(\overline{\varepsilon}_I+\ov p_I)\right]=\frac{3}{c^2}H^2(1+q) =:\frac{1}{\lambda^2}\,
, \ee
where we have defined a new (time-dependent) variable $\lambda$, which has dimension of length.
It is this length scale that will play below the role of the interaction range in the effective Yukawa potential. It is thus of great importance.

Let us now address the perturbations. For simplicity, we restrict
ourselves to scalar perturbations. The metric then reads
\be{7} ds^2= a^2\left[\left(1 + 2\Phi\right)d\eta^2 -\left(1-
2\Psi\right)\delta_{\alpha\beta}dx^{\alpha}dx^{\beta}\right]\, . \ee
It is well known that for the considered perfect fluids we have $\Phi=\Psi$. In what follows, we consider the case of weak gravitational fields, $|\Phi|\ll1$. For
the linearized Einstein equations, one finds the following form \cite{Eingorn1,Eingorn3}:
\be{8} \triangle\Phi-3{\mathcal H}(\Phi'+{\mathcal H}\Phi) =\frac{1}{2}\kappa a^2 \left(\delta T_0^0\right)=\frac{1}{2}\kappa
a^2\left(\frac{c^2}{a^3}\delta\rho_{\rm M}+\frac{3\overline\rho_{\rm M} c^2}{a^3}\Phi +\sum\limits_I\delta{\varepsilon}_I\right)\, , \ee
\ba{9} &{}&\frac{\partial}{\partial x^{\alpha}}(\Phi'+{\mathcal H}\Phi)=\frac{1}{2}\kappa a^2\left(\delta T_{\alpha}^0\right)=-\frac{1}{2}\kappa
a^2\left(\frac{c^2}{a^3}\sum\limits_n\rho_n\tilde v^{\alpha}_n +\sum\limits_I(\varepsilon_I+p_I)\tilde v_I^{\alpha} \right)\, , \ea
\be{10} \Phi''+3{\mathcal H}\Phi'+\left(2{\mathcal H}'+{\mathcal H}^2\right)\Phi=\frac{1}{2}\kappa a^2\sum\limits_I\delta{p}_I\, . \ee
Here, $\triangle$ is the Laplace operator in flat space, and $\tilde v^{\alpha}$ are comoving peculiar velocities. For the case of discrete gravitating masses
$m_n$, one can easily formulate the energy-momentum tensor (see, e.g., equation (106.4) in \cite{Landau}), and one can single out the contribution of the
gravitational potential to the energy density fluctuation: $\delta\varepsilon_{\rm M}\approx \left(c^2/a^3\right)\delta\rho_{\rm M}+\left(3\overline\rho_{\rm M}
c^2/a^3\right)\Phi$, where $\delta\rho_{\rm M}=\rho_{\rm M}-\ov\rho_{\rm M}$. Similarly, for the continuous perfect fluids with energy densities $\varepsilon_I$
and pressures $p_I$, one obtains $\delta\varepsilon_I \approx \delta A_I/a^{3(1+\omega_I)}+3(1+\omega_I){\ov A_I}\Phi/a^{3(1+\omega_I)}$, where the unknown
functions $A_I=\ov{A}_I+\delta A_I$ ($\ov{A}_I=\mathrm{const}$, $\ov\varepsilon_I=\ov{A}_I/a^{3(1+\omega_I)}$) satisfy conservation equations.

On the right-hand sides of \eqref{8}--\eqref{10}, we have inserted the expressions for the fluctuations $\delta T_k^i=T_k^i-\ov{T_k^i}$ of the energy-momentum
tensors for the matter sources \cite{Eingorn1,Eingorn3}. In these expressions, we have taken into account the smallness of the peculiar velocities (at present of
the order of 300 km/s for galaxies) and the smallness of the gravitational potential, but have imposed {\em no} restriction on the fluctuations of the rest mass
density and energy densities (see \cite{Eingorn1,Baumann,EB2} for detailed substantiation). That is, we do not demand the smallness of $\delta\rho_{\rm M}$ and
$\delta A_I$. For this reason, we can use the formalism at all cosmic scales (below and above the Hubble scale).

It is now a straightforward but somewhat lengthy task to derive an equation for the gravitational potential. It was found convenient to split the total
gravitational potential $\Phi$ into individual contributions from each matter source: $\Phi=\Phi_{\rm M}+\sum_I\Phi_I$. Then, these individual gravitational
potentials satisfy the following Helmholtz equations \cite{Eingorn3}:
\ba{28}
\triangle\Phi_{\rm M}-\frac{a^2}{\lambda^2}\Phi_{\rm M}
&=&\frac{\kappa c^2}{2a}\delta\rho_{\rm M}-\frac{3\kappa c^2{\mathcal H}}{2a}\Xi\, ,\\
\label{29} \triangle\Phi_{I}-\frac{a^2}{\lambda^2}\Phi_{I} &=&\frac{\kappa}{2}\frac{\delta
A_I}{a^{1+3\omega_I}}-\frac{3\mathcal{H}\kappa}{2}\frac{1+\omega_I}{a^{1+3\omega_I}}\xi_I\, .
\ea
Here, $\Xi$ and $\xi_I$ are functions that arise from the decomposition of the spatial vectors $\sum_n\rho_n\tilde {\bf v}_n$ and $A_I\tilde {\bf v}_I$ in terms
of gradient and curl.
It is remarkable that all these equations contain {\em the same range} $\lambda$ defined by the formula \rf{5}. What is the value of this range? From observations
we know that the DE equation of state gives $\omega_I\approx-1$ and that the contribution of radiation is currently negligible. Therefore, starting from the
matter dominated epoch, $\lambda$ arises mainly from CDM, and we get from \rf{5}:
\be{26} \lambda \approx \left[\frac{3}{2}\kappa\ov\varepsilon_{\rm
    M}\right]^{-1/2}=\sqrt{\frac{2a^3}{3\kappa\ov\rho_{\rm M} c^2}}=
\sqrt{\frac{2c^2}{9H_0^2\Omega_{\rm M}}\left(\frac{a}{a_0}\right)^3}\,
,\quad \Omega_{\rm M}\equiv \frac{\kappa\ov\rho_{\rm M}
  c^4}{3H_0^2a_0^3}\, . \ee
At the present time ($a=a_0$, $H=H_0$), we obtain $\lambda_0\approx 3700$ Mpc \cite{Eingorn1}.

In the case of discrete sources, equation \rf{28} for the gravitational potential $\Phi_{\rm M}$ can be solved exactly, with the result \cite{Eingorn1,Eingorn3}
\ba{32}
\Phi_{\rm M}&=&\frac{\kappa\overline\rho_{\rm M} c^2\lambda^2}{2a^3}-\frac{\kappa c^2}{8\pi a}\sum\limits_n\frac{m_n}{|{\bf r}-{\bf r}_n|}\exp(-q_n)\nn\\
&+&\frac{3\kappa c^2}{8\pi a}{\mathcal H}\sum\limits_n
\frac{m_n[\tilde{\bf v}_n({\bf r}-{\bf r}_n)]}{|{\bf r}-{\bf
    r}_n|}\,\frac{1-(1+q_n)\exp(-q_n)}{q_n^2}\, ,
\ea
where $q_n(\eta,{\bf r}) := a|{\bf r}-{\bf r}_n|/\lambda=|{\bf R}-{\bf R}_n|/\lambda$ \ (${\bf R}$, ${\bf R}_n$ denote physical distances). Equation~\rf{32}
clearly demonstrates the {\em Yukawa-type exponential screening} of the gravitational potential at distances $|{\bf R}-{\bf R}_n|>\lambda$. We call this effect
the cosmic screening. It is clear from \rf{5} that the cosmic background is responsible for this screening. The resulting equations form the theoretical basis for
subsequent numerical simulations for a very wide class of cosmological models.

The largest known structure in the Universe is the Great GRB Wall (or Hercules-Corona Borealis Great Wall), with the size of the order of $3000\, \mathrm{Mpc}\,
<\lambda_0$ \cite{wall}. It is remarkable that this is consistent with the value for $\lambda_0$.

In the Minkowski background limit, the scale factor $a\to$ const. Then,
$\mathcal{H}\rightarrow0$, and the average energy densities
and pressures tend to zero
(consequently, $\delta\rho_{\rm M}=\rho_{\rm M}-\ov\rho_{\rm M}\to
\rho_{\rm M}$, $\delta A_I=A_I-\ov A_I\to A_I$). We then have $\lambda \to +\infty$ and
\rf{28}, \rf{29} are reduced to the ordinary Poisson equations (the velocity-dependent terms disappear when $\mathcal{H}\rightarrow0$). We also note that the
presence of the exponential functions in \rf{32} makes this sum well defined. In contrast, the well known formulas (8.1) and (8.3) in \cite{Peebles}, which are
derived in the Newtonian approximation, do not have this property.

Another remarkable property of our approach consists in the fact that it predicts the growth of structure in the universe as it should be. It is not difficult to
show from our equations that in the matter dominated stage the density contrast $\delta\rho_{\rm M}/\ov\rho_{\rm M}$ grows linearly with the scale factor:
$\delta\rho_{\rm M}/\ov\rho_{\rm M}\sim a$ for the modes $\lambda_k=a/k <\lambda$ \cite{Eingorn2}.

To summarize, we have shown that general relativity automatically leads to screening of the gravitational interaction at large scales. Our approach incorporates
linear and nonlinear effects with respect to the energy density fluctuations ($\delta\varepsilon/\ov\varepsilon$ can be much larger than $1$). Our results are
consistent with the size of the largest observed structures in the universe. As for potential infinities, they are avoided from the very beginning by the presence
of a Yukawa-type potential. To give Newton the last word \cite{Newton}:
\begin{quote}
You sometimes speak of gravity as essential \& inherent to matter: pray do not ascribe that notion to me, for the cause of gravity is what I do not pretend to
know, \& therefore would take more time to consider of it I fear what I have said of infinites will seem obscure to you: but it is enough if you understand that
infinites when considered absolutely without any restriction or limitation, are neither equal nor unequal nor have any certain proportion to one another, \&
therefore the principle that all infinites are equal is a precarious one.
\end{quote}



\begin{thebibliography}{99}

\bibitem{Harrison} E.~Harrison, {\em Cosmology}, second edition
  (Cambridge University Press, Cambridge, 2000).

\bibitem{Norton} J.~D.~Norton, in: {\em The Expanding Worlds of
    General Relativity}, edited by H.~Goenner {\em et al.}
  (Birkh\"auser, Boston, 1999), p.~271.

\bibitem{Seeliger} H.~Seeliger, {\em Astronomische Nachrichten} {\bf
    137}, 129 (1895).

\bibitem{Mukhanov-book}
V. Mukhanov, {\em Physical Foundations of Cosmology} (Cambridge University Press, Cambridge, 2005).


\bibitem{Eingorn1}
M. Eingorn, {\em The Astrophysical Journal} {\bf 825}, 84 (2016); arXiv:1509.03835 [gr-qc].

\bibitem{Eingorn3} M.~Eingorn, C.~Kiefer, and A.~Zhuk, {\em Journal of
    Cosmology and Astroparticle Physics} {\bf 09}, 032 (2016); arXiv:1607.03394 [gr-qc].

\bibitem{Landau}
L.~D. Landau and E.~M. Lifshitz,
 {\em The Classical Theory of Fields}, fourth edition, volume~2
 (Oxford Pergamon Press, Oxford, 2000).



\bibitem{Baumann}
D. Baumann, A. Nicolis, L. Senatore, and M. Zaldarriaga, {\em Journal of Cosmology and Astroparticle Physics} {\bf 07}, 051 (2012); arXiv:1004.2488 [astro-ph.CO].

\bibitem{EB2}
R. Brilenkov and M. Eingorn, 
{\em The Astrophysical Journal} {\bf 845}, 153 (2017); arXiv:1703.10282 [gr-qc].



\bibitem{wall} I.~Horvath, Z.~Bagoly, J.~Hakkila, and L.~V.~Toth,
{\em Astronomy and Astrophysics} {\bf 584}, A48 (2015); arXiv:1510.01933 [astro-ph.HE].

\bibitem{Peebles}
P. J. E. Peebles, {\em The large-scale structure of the Universe} (Princeton University Press, Princeton, 1980).

\bibitem{Eingorn2}
M. Eingorn and R. Brilenkov, 
{\em Physics of the Dark Universe} {\bf 17}, 63 (2017); arXiv:1509.08181 [gr-qc].


\bibitem{Newton} I.~Newton, Letter to R.~Bentley, January~17, 1693,
http://www.newtonproject.ox.ac.uk.

\end{thebibliography}
\end{document}